\documentclass[12pt,french,a4paper]{amsart}
\usepackage[T1]{fontenc}

\input{amssym.def}
\newcommand{\gid}[1]{\langle #1 \rangle}

\newcommand{\intervalle}[2]{[ \negthinspace [ #1 , #2 ] \negthinspace ]}

\newcommand{\calr}{{\mathcal{R}}}
\newcommand{\MFM}{{\mathfrak{M}}}
\newcommand{\MFJ}{{\mathfrak{J}}}
\newcommand{\MFP}{{\mathfrak{P}}}
\newcommand{\MFG}{{\mathfrak{G}}}

\newcommand{\Psym}{{\Psi_{sym}}}
\newcommand{\Qsym}{{\Theta_{sym}}}

\newcommand{\Salpha}{{\underline{\alpha}}}
\newcommand{\Sbeta}{{\underline{\beta}}}

\newcommand{\Card}{\operatorname{Card}}

\font\twelvebb=msbm10 at 11pt

 \def\Q{\hbox{\twelvebb Q}}

\newtheorem{bidon}{Bidon}

\newtheorem{propo}[bidon]{Proposition}
\newtheorem{corollaire}[bidon]{ Corollaire}
\newtheorem{theorem}[bidon]{{Th\'eor\`eme}}
\newtheorem{lemme}[bidon]{{Lemme}}

\theoremstyle{remark}

\newtheorem{rem}[bidon]{{\bf Remarque}}

\newtheorem{ex}[bidon]{{\bf Exemple}}
\newtheorem{exs}[bidon]{{\bf Exemples}}
\newtheorem{notation}[bidon]{{\bf Notation}}
\newtheorem{notation-def}[bidon]{{\bf Notation-Définition}}
\newtheorem{note}[bidon]{{\bf Note}}
\newtheorem{conv}[bidon]{{\bf Convention}}

\begin{document}
\newenvironment{resfr}{
      \begin{center}
        \bfseries Résumé
     \end{center}
}

\newenvironment{reseng}{
      \begin{center}
        \bfseries Abstract
      \end{center}
}


\title{La Résolvante de Lagrange et ses Applications}
\author{Annick Valibouze }
\address{
L.I.P.6, Université Pierre et Marie Curie,
4, place Jussieu,
F-75252 Paris Cedex 05
}
\email{annick.valibouze@upmc.fr}
\date{\today}

\subjclass[2000]{Primary 12F10; Secondary 12Y05, 11Y40}
\keywords{Groupe de Galois, Résolvantes, Matrices de Partitions et de
  Groupes}

\maketitle

\bigskip
\begin{resfr}
Dans cet article, les changements de représentations d'un groupe sont utilisés
pour décrire son action en tant que groupe 
de Galois d'un polynôme sur les racines des facteurs
simples d'une quelconque 
de ses résolvantes de Lagrange. Ainsi est déterminé le groupe de
Galois de la
résolvante mais aussi celui de chacun de ses facteurs. Nous exposons
ensuite différentes applications.
En particulier, par ce biais, sont retrouvés des
résultats classiques de la théorie de Galois constructive.

\end{resfr}

\begin{reseng}
This paper describes the action of the Galois group of a univariate
polynomial on the factors of any of its 
resolvents. 
\end{reseng}

\section*{Introduction}

En introduisant la résolvante J.L. Lagrange (voir
\cite{Lagrange:1770}) unifia les résultats obtenus par ses
prédécesseur pour résoudre les équations jusqu'au quatrième degré.
Avec ses résolvantes, prélude aux célèbres sommes de Gauss, il
introduisit les groupes de permutations dans la résolution des équations 
algébriques. L'idée de J.L. Lagrange est de faire agir un sous-groupe
$L$ du groupe symétrique sur un polynôme $r$
de plusieurs variables et d'observer ce qui se passe quand ces variables se
spécialisent en les solutions de l'équation.
 Plus tard, E. Galois identifia le groupe de l'équation comme celui
 échangeant les racines du polynôme minimal d'un élément primitif du
 corps des solutions de l'équation ; il fit agir ce groupe
 sur les spécialisations~;
 cette façon d'étudier le groupe de l'équation, appelé aujourd'hui
 Groupe de Galois, restreint le champs d'investigations lorsqu'il
 s'agit de le déterminer. En effet, si une permutation n'appartient pas au
 groupe de Galois, l'action n'est pas définie (voir Paragraphe
 \ref{section : auto}) et, a priori, seule l'identité appartient
 de façon certaine au groupe de Galois. Par la suite, les travaux
 d'E. Artin permirent d'énoncer la correspondance galoisienne. Si ce point
 de vue apporte une vision théorique fructueuse et utile, il reste
 difficile de mener des calculs dans le corps des racines avec un
 groupe de $k$-automorphismes non identifiés a priori.

Pour la
détermination du groupe de Galois d'un polynôme et de son corps des
racines, le point de vue de J.L. Lagrange est le plus fructueux. Le
polynôme $r$ sur lequel agissent les permutations de $L$ est un
invariant (précisément un invariant $L$-primitif)
d'un sous-groupe $H$ de $L$. 
Par conséquent, il est possible de s'affranchir du polynôme
$r$ pour ne réaliser que des pré-calculs groupistiques.

Dans cet article, cette démarche groupistique est poussée jusqu'au
point de pré-déterminer les groupes de Galois des résolvantes (et donc de
leurs facteurs) d'un
polynôme d'une variable par de
simples changements de représentations du groupe de Galois de ce
polynôme. Il aboutit aux matrices de groupes (déterminées différemment 
dans \cite{Valibouze:95}).

Ce travail s'inscrit dans la suite des travaux d'E.H. Berwick
(voir \cite{berwick:29}), de Foulkes (voir \cite{Foulkes:31}) et de
ceux, plus récents, de J. Mc Kay et G. Butler (voir 
\cite{ButlerMcKay:83}) et de bien d'autres. Il
reprend et complète l'article
\cite{ArnaudiesValibouze:97} aboutissant aux matrices
dites de partitions (détermination des degrés des facteurs
des résolvantes).

Cet article décrit clairement la composante résolvante de la
théorie de Galois constructive. 
Seule la définition classique
de la résolvante 
est considérée car ne sont abordés ni son aspect
calculatoire ni celui de la détermination du
groupe de Galois obtenue simultanément au calcul du corps de décomposition du
polynôme (voir \cite{Ducos:2000}, \cite{Valibouze:99} et
\cite{Valibouze:05}). Via le livre de 
N. Tchebotarev (voir 
\cite{Tchebotarev:50}), le lecteur pourra pousser plus avant l'étude
de la 
théorie de Galois constructive du point de vue des idéaux poursuivant ainsi
celui de J.L. Lagrange. 

Afin que cet article soit abordable par les non spécialistes, les trois
 premiers paragraphes sont dévolus à une introduction
 rapide à la théorie de Galois unifiant différentes approches : idéal
 des relations, groupe de Galois 
 en tant que groupe de permutations et en tant que groupe des
 $k$-automorphismes du corps des racines, correspondance
 galoisienne. Certaines nouvelles démonstrations de théorèmes connus
 y sont proposées. Les matrices de groupes sont définies au quatrième
 paragraphe. Les paragraphes 5 à 9 sont consacrés à la résolvante,
 son groupe de Galois, les groupes de Galois de ses facteurs, le corps
 de ses racines. Le dixième paragraphe illustre les résultats avec des
 résolvantes connues. Le dernier paragraphe est consacré aux applications.

\section*{Données et notations préliminaires}

Nous fixons 
$n$ variables $x_1,x_2,\ldots ,x_n$ algébriquement indépendantes.
Soit $\alpha_1,\ldots , \alpha_n$ une numérotation des $n$ racines d'un
polynôme 
$f$ de degré $n$ à coefficients dans un corps parfait $k$. Posons $\Salpha=(\alpha_1,\ldots ,
\alpha_n)$.

Le corps des racines du polynôme $f$ est noté
$k(\Salpha)$. Ce corps est la plus petite extension
algébrique de $k$ dans lequel le polynôme $f$ se factorise
entièrement en facteurs linéaires ; ce qui fait qu'il s'appelle aussi corps de
décomposition de $f$.

Le groupe des permutations d'un ensemble $E$ est noté $S_E$ et si $E$
est l'ensemble $\{1,2,\ldots ,n\}$ alors $S_E$ est le groupe symétrique de degré $n$, noté aussi
$S_n$.
Ce groupe agit naturellement sur les polynômes de
$k[x_1,\ldots ,x_n]$ par permutations des indices des variables :
$\sigma.p=p(x_{\sigma(1)},\ldots , x_{\sigma(n)})$ pour
$p \in k[x_1,\ldots ,x_n]$ et $\sigma \in S_n$.

\section{Idéal des relations, groupe de Galois et corps des racines}

L'{\it idéal $\MFM$ des $\Salpha$-relations (sur $k$)} est défini par~:
$$
\MFM= \{ r \in k[x_1,\ldots ,x_n] \;\mid \; r(\Salpha)=0 \} 
$$
et considérons l'anneau quotient~:
 $$K_\Salpha=k[x_1,\ldots ,x_n]/\MFM \quad .$$ 
Nous constatons que $\MFM$ est défini en observant l'évaluation
d'un polynôme $r$ en les racines du polynôme $f$. C'est donc le point de
vue de J.L. Lagrange qui s'applique.

Le {\it groupe de 
Galois $G$ de $\Salpha$ sur $k$} se définit comme le sous-groupe du
groupe symétrique 
$S_n$ stabilisant globalement l'idéal $\MFM$~:
$$
G = \{\sigma \in S_n \;\mid \; \sigma.\MFM =\MFM \} \; .
$$
où $\sigma.\MFM$ est l'ensemble des permutés $\sigma.r$ où $r$
parcourt $\MFM$. 

\begin{rem} \label{rem : sens} Ici, nous touchons le point clef qui donne la préférence
 au point de vue de J.L. Lagrange. 
Sans erreurs et sans connaître $G$ a priori, il est possible
 de faire agir toute 
permutation du groupe symétrique $S_n$ car il s'agit de polynômes
génériques sur lesquels l'action est définie. En effet,
nommons $\alpha_1=1,\alpha_2=j$ et $\alpha_3=j^2$ les racines du polynôme 
$x^3-1$ et choisissons la permutation $\sigma=(1,2)$. Nous avons l' $\Salpha$-relation
$x_2^2-x_3$ inexistante pour un polynôme générique.
A quoi correspondrait $\sigma.(\alpha_2^2)$ ?, à $\alpha_1^2=1$ ou bien
à $\sigma.\alpha_3=\alpha_3$ ? L'action n'est donc pas définie si
$\sigma$ n'appartient pas à $G$ qui est précisément le plus grand
sous-ensemble
de $S_n$ pour lequel l'action ait un sens.
\end{rem}

Par le $k$-morphisme d'évaluation de l'anneau $k[x_1,\ldots ,x_n]$
dans le corps
$k(\Salpha)$
qui à $x_i$ associe $\alpha_i$, de noyau
$\MFM$, l'anneau
quotient $K_\Salpha$ est isomorphe au corps $k(\Salpha)$. 

La dimension
$\text{dim}_k(k(\Salpha))$
du corps $k(\Salpha)$ en tant que $k$-espace vectoriel, appelée aussi degré
de l'extension $k(\Salpha)/k$, satisfait l'identité~:

\begin{eqnarray}
\text{dim}_k(k(\Salpha)) =\text{Card}(G) \quad \label{eqn : dim}.
\end{eqnarray}

En effet, le $k$-isomorphisme entre les corps $K_\Salpha$ et
$k(\Salpha)$ induit l'égalité~:  
$$
\text{dim}_k(k(\Salpha)) = \text{dim}_k(K_\Salpha)\quad .
$$
Or la dimension $\text{dim}_k(K_\Salpha)$ est
identique au cardinal de la variété $V$ de $\MFM$ puisque cet idéal est
radical (il est maximal puisque $k(\Salpha)$ est un corps).
La variété $V$ est l'ensemble des $(\alpha_{\sigma(1)},\ldots
,\alpha_{\sigma(n)})$ où $\sigma$ parcourt $G$ (voir
\cite{Valibouze:99}). Comme les racines de $f$ 
sont distinctes deux-à-deux, le cardinal de $V$ est identique à celui
du groupe de Galois $G$.

\section{Groupe de Galois et groupe des $k$-automorphismes}
\label{section : auto}

Soit $E$ une $k$-algèbre. Un $k$-endomorphisme de $E$ (en tant que
$k$-algèbre) est une
application $\phi$ de $E$ dans $E$ telle que si $e_1,e_2 \in E$ et
$\lambda \in k$ alors $\phi(\lambda)=\lambda$,
$\phi(e_1e_2)=\phi(e_1)\phi(e_2)$ et $\phi(e_1+e_2)=\phi(e_1)+ \phi(e_2)$.
Si $\phi$ est surjectif alors $\phi$ est un $k$-automorphisme. L'ensemble de
$k$-automorphismes de $E$ est le groupe noté Aut$_k(E)$.\\

Chaque $k$-endomorphisme de $k(\Salpha)$ laissant invariants les
coefficients de $f$, il est induit par une permutation de ses
racines. 
Donc tout $k$-endomorphisme de $k(\Salpha)$ est un
$k$-automorphisme et
nous pouvons définir une
représentation, dite {\it associée à
$\Salpha$}, du groupe $Aut_k(k(\Salpha))$ dans $S_n$~:
$$
\begin{array}{lccl}
 & Aut_k(k(\Salpha)) &\longrightarrow &  S_n \\
&\phi & \mapsto & \sigma_\phi : \alpha_{\sigma_\phi(i)}= \phi(\alpha_i) \quad .
\end{array}
$$

Notons $\mathcal F$ le $k$-isomorphisme du corps $K_\Salpha$ dans le corps 
$k(\Salpha)$ qui à $p$ associe $p(\Salpha)$.

\begin{lemme} Soit $\sigma \in G$.  Soit le $k$-endomorphisme de
permutations $\mathcal G_\sigma$ de $K_\Salpha$ qui à $p$ associe
$\sigma.p$. Alors $\mathcal G_\sigma$ est un 
$k$-automorphisme. Par conséquent, l'application
$$\phi_\sigma=\mathcal{ F G_\sigma F}^{-1}$$
est un $k$-automorphisme de $k(\alpha_1,\ldots
,\alpha_n)$ satisfaisant 
$$\phi_\sigma(i)=\alpha_{\sigma(i)}$$ pour tout $i\in
\intervalle{1}{n}$.
\end{lemme}

\begin{proof}
Car, par définition, la condition $\sigma \in G$ est
équivalente à $\sigma.\MFM=\MFM$.
\end{proof}

\begin{lemme} Le groupe $G$ est la représentation de
  $Aut_k(k(\alpha_1,\ldots 
,\alpha_n))$ dans $S_n$ associée à $\Salpha$ et l'image réciproque de $\sigma
\in G$ par cette représentation est le $k$-automorphisme
$\phi_\sigma$. \\
Soient $p \in k(\Salpha)$ et $P={\mathcal
  F}^{-1}(p)$ appartenant à
$K_\Salpha$. Alors pour tout $\sigma \in G$~:
$$
\phi_\sigma(p) = (\sigma.P)(\alpha_1,\ldots ,\alpha_n) \quad .
$$
\end{lemme}
\begin{proof}
Soient $\phi \in Aut_k(k(\alpha_1,\ldots 
,\alpha_n))$ et $R \in k[x_1,\ldots ,x_n]$ tel que $r=R(\alpha_1,\ldots
,\alpha_n)=0$. Posons $\sigma =\sigma_\phi$. Nous avons
$$(\sigma.R)(\alpha_1,\ldots ,\alpha_n)=R(\alpha_{\sigma(1)},\ldots
,\alpha_{\sigma(n)})=R(\phi(\alpha_1),\ldots ,\phi(\alpha_n))=\phi(r)=0$$ car
$\phi$ est un $k$-automorphisme. Donc $\sigma \in G$.
\end{proof}

\begin{notation} \label{nota : essentiel}
D'après les deux lemmes précédents et ayant fixé la numérotation des racines de $f$, 
pour tout $\beta \in k(\alpha_1,\ldots ,\alpha_n)$ et tout $\sigma \in G$,
nous pouvons poser~:
$$
\beta^\sigma = \phi_\sigma(\beta) \quad .
$$
\end{notation}

\begin{rem}\label{rem : sens2} En appliquant la notation 
  \ref{nota : essentiel}, $G$ est le 
plus grand sous-groupe de $S_n$ assurant que pour tout $\sigma \in G$ et
 $\gamma \in k(\Salpha)$ si $\gamma=0$ alors $\gamma^\sigma=0$. La notation
\ref{nota : essentiel} n'a de sens que pour $\sigma \in G$. En 
revanche, pour $P\in k[x_1,\ldots ,x_n]$ et $\sigma \in S_n$, la notation
$(\sigma.P)(\Salpha)$ en a un (voir Remarque
\ref{rem : sens}).
\end{rem}

\begin{conv} Lorsque nous voudrons désigner une représentation symétrique
quelconque de $Aut_k(k(\alpha_1,\ldots ,\alpha_n))$ dans $S_n$, nous
l'appellerons {\it groupe de Galois de $f$ sur $k$} et nous la
noterons Gal$_k(f)$.  
\end{conv}

{\bf Note} Le groupe Aut$_k(k(\Salpha))$ est aussi communément appelé le groupe de
Galois de l'extension $k(\Salpha)/k$. Par abus de
langage, le groupe $G$ (et donc aussi Gal$_k(f)$) est souvent appelé le
groupe de Galois de cette extension.

\section{La correspondance galoisienne}
\label{section : corgal}
Le polynôme minimal sur $k$ de tout
$\beta$ appartenant à $k(\alpha_1,\ldots
,\alpha_n)$ est donné par~:
\begin{eqnarray} \label{eqn : minimal1}
Min_{\beta,k}=\prod_{\gamma \in \{\beta^g \; \mid \; g \in G\}}(x-\gamma)
\end{eqnarray}
Cette identité est démontrable par de l'algèbre linéaire sur le
$k$-espace vectoriel $K_\Salpha$ (voir \cite{Valibouze:99}).\\

D'après le théorème de l'élément primitif de
J.L. Lagrange (voir Note \ref{note : prim}), 
il existe $v \in k(\Salpha)$ tel 
que 
$$k(\Salpha)=k(v)\quad .$$
Cet élément s'exprime sous la forme 
$$v=V(\alpha_1,
\alpha_2,\ldots ,\alpha_n)$$ 
où $V\in k[x_1,\ldots ,x_n]$. Le degré du polynôme
minimal de $v$ sur $k$ est $d$, l'ordre du groupe de Galois $G$
(puisque, d'après l'identité (\ref{eqn : dim}), c'est le degré de
l'extension $k(\Salpha)/k$). 

\begin{note}\label{note : geneideal}
L'idéal $\MFM$ est calculable à partir de l'idéal $\MFJ$
des relations symétriques (voir \cite{LITP9361}) :
$$
\MFM = \MFJ \; + \, \gid{Min_{v,k}(V)} \qquad .
$$
Mais cela nécessiterait d'abord d'obtenir Min$_{v,k}$ par le calcul
et la factorisation du polynôme 
$$R_V= \prod_{\sigma \in S_n}(x- (\sigma.V)(\Salpha))$$
de degré $n!$ et
d'ensuite de calculer l'ensemble triangulaire engendrant $\MFM$. Ce
dernier calcul peut s'avérer très complexe si l'ordre du groupe de
Galois
est élevé. Le lecteur pourra consulter les articles
\cite{Ducos:2000}, \cite{Sargov:2003.05},
\cite{Valibouze:99} et \cite{Valibouze:05} présentant des méthodes
plus efficaces pour le calcul de $\MFM$.
\end{note}

{\bf Note} L'historique résolvante dite de Galois
est le polynôme 
$$ \text{Min}_{v,k}=\prod_{g \in G}(x-v^g) $$
ou bien tout autre facteur sur $k[x]$ du polynôme $R_V$
(c'est
selon selon
les auteurs). E. Galois définit le groupe de l'équation comme celui
échangeant les racines de Min$_{v,k}$. L'approche proposée ici est de le
définir comme le groupe stabilisant l'idéal des relations et
d'aboutir ensuite à la formule (\ref{eqn : minimal1}). Le théorème qui
suit est connu sous le nom de {\it Théorème de Galois}.

\begin{theorem} (\cite{Galois:97}) \label{theo : galois}
Soit $\beta \in k(\Salpha)$. Pour que
$\beta$ appartienne à $k$ il faut et il suffit que $\beta^g=\beta$ pour tout
$g \in G$.
\end{theorem}

\begin{proof} 
Soit $P \in k[x]$ de degré au plus $d-1$ tel que 
$\beta =P(v)$.
Le polynôme 
$$ W(x) =  P(x) - \beta $$
de degré strictement inférieur à $d$ appartient à $k(\Salpha)[x]$. De
plus, pour chaque $g \in G$, 
l'identité $\beta^g=\beta$ est équivalente à 
$W(v^g)=0$ (voir Notation
\ref{nota : essentiel}). \\  
Si $\beta^g=\beta$ pour tout $g \in G$ alors les $d$ racines distinctes
du polynôme minimal de $v$ sur $k$ sont aussi racines de $W$ de degré $d-1$.
Le polynôme $W$ est donc nul et par suite $\beta = P(0) \in k$.
Inversement, si $\beta \in k$ alors $W(V)=P(V) - \beta$ appartient à l'idéal
des 
relations $\MFM$ car $V(\Salpha)=v$. Donc, par définition de $G$, $W(v^g)=0$ pour tout $g \in G$. Ce qui termine la démonstration.
\end{proof}

{\bf Note} La dernière partie de cette démonstration est un bon reflet
de la différence entre l'approche lagrangienne
que nous adoptons et l'approche galoisienne. Comme E. Galois,
introduisons le groupe $\mathcal G$ échangeant les racines du
polynôme Min$_{v,k}$. Pour montrer que $W(v^g)=0$ pour tout
$g \in \mathcal G$, il y a deux solutions. La
première consiste à remarquer que $W$ possède une racine en commun avec
Min$_{v,k}$
et de déduire de l'irréductibilité de ce dernier que $W$ possède toutes
ses racines $v^\sigma$, $\sigma \in \mathcal G$ ; ce
qui nécessite la démonstration d'un lemme préalable. La seconde est de 
chercher à faire agir $\mathcal G$ sur 
$P(v)-\beta=0$. C'est ce que font de nombreux auteurs avec beaucoup de
contorsions, voir avec des erreurs~; cette deuxième solution fonctionne
parce que 
$\mathcal G=G$ (voir Remarque \ref{rem : sens2}).

Une extension de $k$ est dite {\it galoisienne} si elle est le corps
des racines d'un polynôme de $k[x]$.
 
\begin{notation} Soit $H$ un sous-groupe du groupe de Galois $G$. La notation
$k(\Salpha)^H$ désigne le sous-corps de $k(\Salpha)$ formé de ses éléments
$\beta$ tels que $\beta^\sigma \in k(\Salpha)^H$ pour tout $\sigma \in H$ (
i.e. invariants par toute permutation de $H$).
\end{notation}

Le théorème \ref{theo : galois} s'exprime sous la forme~:
$$
k=k(\Salpha)^G \quad .
$$
L'identité $k(\Salpha)=k(\Salpha)^{I_n}$, où $I_n$ est le sous-groupe
identité de $S_n$, conduit à s'interroger sur le lien existant entre
les sous-groupes de $G$ et les corps
intermédiaires entre $k$ et $k(\Salpha)$.
C'est la correspondance galoisienne qui y répond. Elle s'exprime en les
points suivants~:\\

1. Si $K$ est un corps intermédiaire entre $k$ et $k(\Salpha)$ alors il existe
un sous-groupe de $G$ tel que $K=k(\Salpha)^H$. \\

2. Si $H$ est un sous-groupe de $G$ alors il existe un sous-corps $K$ de
   $k(\Salpha)$ tel que $K=k(\Salpha)^H$.\\

3. Dans chacun de ces cas, d'après le Lemme d'Artin, l'extension
$k(\Salpha)$ de $K$ est galoisienne et
le groupe $H$ est le groupe de
Galois de $\Salpha$ sur $K$ ; l'extension $k(\Salpha)$ de $K$ est
donc de degré l'ordre du groupe 
$H$ et l'extension $K$ de $k$ est de degré l'indice de $H$ dans $G$ ; si, de
plus, $H$
est un sous-groupe distingué de $G$ alors l'extension $K/k$ est
galoisienne et le groupe quotient $G/H$ est isomorphe au
groupe des $k$-automorphismes de $K$.

\begin{propo} \label{propo : minbeta}
Soient $H_1\subset H_2$ deux sous-groupes de $G$ et
  $\beta \in k(\Salpha)$.
L'égalité 
$$\text{Stab}_{H_2}(\beta) =H_1$$
est satisfaite si et
seulement si $\beta$ est un élément primitif du corps
$k(\Salpha)^{H_1}$ sur le corps $k(\Salpha)^{H_2}$ et dans ce cas
$$
Min_{\beta,k} = \prod_{\sigma \in H_2/H_1}(x- \beta^\sigma) \qquad .
$$
\end{propo}
\begin{proof} Comme l'impliquent les deux assertions de la proposition,
  nous avons $\beta \in k(\Salpha)^{H_1}$.
Le polynôme minimal de $\beta$ sur $k(\Salpha)^{H_2}$ est de
degré au plus $d$, l'indice de $H_1$ dans $H_2$ (i.e. le degré de l'extension).
$H_2$ étant le groupe de Galois
de $f$ sur $k(\Salpha)^{H_2}$, les racines de ce polynôme sont les
$\beta^\sigma$ où $\sigma$ parcourt $H_2$.\\
 Si Stab$_{H_2}(\beta)
=H_1$, il existe exactement $d$ racines distinctes :
celles obtenues en parcourant $H_2/H_1$. Le
polynôme minimal de $\beta$ étant de même degré que
l'extension considérée, $\beta$ est un élément primitif de
cette extension. Inversement s'il existait $\sigma \in H_2\backslash H_1$ tel
que $\beta^\sigma=\beta$ alors le polynôme minimal de $\beta$
serait de degré strictement inférieur à $d$ ; ce qui
contredirait la primitivité de $\beta$.
\end{proof}
\section{Matrices des groupes et des partitions}
\label{para : 1}
Soit $L$ un sous-groupe de $S_n$ et $G$ et $H$ deux sous-groupes de $L$.
Nous notons $e$ l'indice de $H$ dans $L$.
Nous faisons agir $G$ \`a gauche sur $L/H$, les classes
\`a gauche de $L$ modulo $H$. Nous définissons ainsi une représentation
(naturelle) par permutations de $G$ dans le groupe $S_{L/H}$ :
$$ \begin{array}{lccc}
\Psi :& G &\longrightarrow &S_{L/H} \\
       & g &\mapsto &\sigma_g 
\end{array}
$$
telle que, pour $C,C^\prime \in {L/H}$, 
$\sigma_g.C = C^\prime$ si $gC =  C^\prime$.
 Par $\mathcal O$, nous désignons
l'ensemble des orbites pour cette représentation.\\

\begin{note} \label{note : multiso} Soient $g,g^\prime \in G$. L'identité
$\Psi(g)=\Psi(g^\prime)$ est satisfaite si et seulement si $g^\prime \in gJ$ où
$$J=\bigcap_{\sigma \in L}H^\sigma$$ est
  un sous-groupe normal de $L$.
Le groupe $G$ est $m$-isomorphe au groupe $\Psi(G)$ où $m$ est l'ordre
du groupe $N=J\cap G$. Le groupe $G/N$ est simplement isomorphe à 
$\Psi(G)$. Si $H \not \in \{S_n,A_n,V_4,D_4\}$ alors $N$ est le groupe
identité. 
\end{note}

\begin{notation} La notation 
$$P_L(G,H) \quad ,$$
 ou plus simplement $P(G,H)$,
désignera la  
partition $1^{m_1},2^{m_2},...,e^{m_e}$ o\`u, 
pour $i\in \intervalle{1}{e}$, l'entier $m_i$ est
le nombre d'orbites de cardinal $i$
(nous retirons les $i^0$ de $P(G,H)$ et posons $i=i^1$) par action de
$\Psi(G)$
sur $L/H$. Nous avons
$e= m_1+2m_2+\cdots +em_e$, le poids de la partition, et
nous posons $m=m_1+\cdots + m_e$ sa longueur qui est
le nombre d'orbites (i.e. le cardinal de
$\mathcal O$).
\end{notation} 
 
\begin{exs} \label{ex : 6.3}
Ces exemples seront poursuivis pour illustrer les résultats essentiels.

{\bf 1.} Pour $L=S_4$, $G=D_4=\gid{(1,2,3,4), (1,3)}$, un groupe diédral dans $S_4$ et $H=A_4$, le groupe alterné, nous avons
$$S_4/A_4=\{A_4,(3,4)A_4\} \quad , \mathcal O=\{\{A_4,(3,4)A_4\}\} \,
\text{ et } \, P_{S_4}(D_4,A_4)=2 \; .$$ 

{\bf 2.} Pour $L=S_4$, $G=D_4$ et $H=S_2\times S_2$, nous
avons $P_{S_4}(D_4,H)=2,4$ avec
$$\mathcal O=\{\{(2,3)H,(1,2,4,3)H\},\{H,(1,2,3)H,(1,3)(2,4)H,(2,4,3)H\}\}
\quad .$$

{\bf 3.} Pour $L=M_5=\gid{(1,2,3,4,5), (1,2,4,3)}$, le groupe méta-cyclique de
degré 
5, $G=C_5=\gid{(1,2,3,4,5)}$ et $H=D_5=\gid{(1,2,3,4,5), (2,5)(3,4)}$, nous
avons 
$$\mathcal O=\{\{D_5\},\{(2,3,5,4)D_5\}\}
\quad \text{ et } \quad P_{M_5}(C_5,D_5)=1^2 \quad .$$
De même, soient $H_1=Id_5$, $H_2=\gid{(2,5)(3,4)}$,
$H_3=\gid{(2,5)(3,4), (2,3,5,4)}$, $H_4=C_5$, $H_5=D_5$ et $H_6=M_5$ des
représentants des six classes de conjugaisons dans $M_5$. La matrice 
$\MFP=(P(H_i,H_j))_{1\leq i,j \leq 6}$ est la suivante~:
$$
\MFP= \left( \begin{array}{cccccc}
      1^{20}& 1^{10}   & 1^{5} & 1^4 & 1^2 & 1 \\
      2^{10} & 1^2,2^4 & 1,2^2 & 2^2 & 1^2 & 1\\
      4^5    & 2,4^2   & 1,4   & 4   & 2  & 1\\
      5^4    & 5^2  &     5    & 1^4  & 2 & 1\\
      10^2   & 5^2  &     5    & 2^2  & 1^2 & 1\\
      20     & 10   &     5    &  4   &  2  & 1 
\end{array}\right ) \quad .
 $$
Les partitions d'une même colonne $j$ ont comme poids l'indice
de $H_j$ dans le groupe $M_5$.
Nous verrons plus loin que cette matrice ne dépend pas des représentants
choisis pour chaque classe de conjugaison.
\end{exs}
 
La représentation $\Psi$ de $G$ dans $S_{L/H}$ est équivalente à une
représentation symétrique $\Psym$  
de $G$ dans $S_e$ induite par un ordre sur les $e$ classes de $L/H$~:
$$
\Psym \; : \; G \longrightarrow S_{L/H} \longrightarrow S_e
$$
\begin{conv}\label{conv : 6.4}
Afin de simplifier la présentation,
nous choisissons d'ordonner les
classes de $L/H$ de telle manière que les classes d'une même orbite soient
consécutives et que les classes d'une orbites de cardinal $c$ soient ordonnées
avant celles des orbites de cardinal supérieur à $c$ (ce n'est pas un ordre
total).
\end{conv}

Le produit direct de groupes symétriques $S_1^{m_1}\times S_2^{m_2} \times
\cdots \times S_e^{m_e}$ est usuellement noté
$S_{1^{m_1},2^{m_2},...,e^{m_e}}$. Avec la convention que nous avons choisie,
la représentation symétrique $\Psym(G)$
de $G$ dans $S_e$ est un sous-groupe de $S_{P(G,H)}$. \\  

\begin{notation-def}\label{nota : 6.6} 
Soit $j\in \intervalle{1}{m}$.
Notons $p_j$ le cardinal de la $j$-ième orbite $O$.
Le groupe noté 
$$\Psi(G)_j$$ 
est une représentation symétrique transitive de $G$
dans $S_{p_j}$ induite par une représentation de $G$ par action à
gauche sur l'orbite $O$.
La notation $$Gr_L(G,H) \quad ,$$
ou plus
simplement  $Gr(G,H)$, désignera la suite $\Psi(G)_1,\ldots ,
\Psi(G)_m$.
\end{notation-def}

\begin{rem} \label{rem : 6.5} 
Pour $j\in \intervalle{1}{m}$, 
$$\Psi(G)_j$$ est aussi le sous-groupe de $S_{p_j}$
obtenu par l'action
du groupe de permutations $\Psym(G)$ sur l'ensemble des $p_j$ entiers
$\{p_1+\cdots +p_{j-1}+1,\ldots ,p_1+p_2+\cdots +p_j\}$ (en posant
$p_0=0$).
\end{rem}



Reprenons nos exemples, en choisissant chaque fois un ordre sur les orbites.

\begin{exs}\label{ex : 6.7}{\ }  \\
{\bf 1. (suite)}. Nous avons $Gr_{S_4}(D_4,A_4)=S_2$
car $P_{S_4}(D_4,A_4)=2$.\\

{\bf 2. (suite)}. Nous avons $P_{S_4}(D_4,S_2\times S_2)=2,4$. L'action de
$D_4$ sur l'orbite de 
cardinal 4 montre que $Gr_{S_4}(D_4,S_2\times S_2)=S_2, D_4$.\\

{\bf 3. (suite)} Nous avons $Gr_{M_5}(C_5,D_5)=S_1,S_1= S_{1}^2$ car
$P_{M_5}(C_5,D_5)=1^2$.
\end{exs}

{\bf Note} La démonstration de la proposition suivant adapte celle de la
proposition 10 de
\cite{ArnaudiesValibouze:97} portant sur la partition $P_L(G,H)$.

\begin{propo} \label{propo : 6.8} La suite
  $Gr_L(G,H)$ ne dépend que des classes de conjugaison 
de $G$ et $H$ dans $L$.
\end{propo}

\begin{proof} 
Posons $G^\tau=\tau G \tau^{-1}$ et $H^\tau = \tau H
  \tau^{-1}$, $\tau \in L$. Nous avons la bijection naturelle :
\(
\begin{array}{lccc}
h :& L/H &\longrightarrow& L/H^\tau\\
   & C   & \mapsto & C \tau^{-1}
\end{array}
\) .

Soit $\Psi^{\prime}$ la représentation de $G$ dans $S_{L/H^\tau}$. La
suite
$Gr_L(G,H)$ ne dépend que de la classe de conjugaison
de $H$ dans $L$ car
les représentations $\Psi$ et $\Psi^{\prime}$ sont équivalentes. En effet,
pour tout $C \in  L/H$, $g \in G$, en posant 
$C^\prime = gC=\Psi(g).C$, nous avons 
$$h^{-1}o\Psi^{\prime}(g)oh(C) =  h^{-1}o\Psi^{\prime}(g)(C\tau^{-1})=
h^{-1}(gC\tau^{-1})= 
h^{-1}(C^\prime\tau^{-1})=C^\prime=\Psi(g).C \quad .$$

Montrons l'indépendance du choix de $G$ dans sa classe d'équivalence. Soit
$I_\tau$ l'automorphisme de conjugaison de $L$ dans $L$ qui à $\sigma$ associe
$\sigma^\tau$. L'ensemble $L/H^\tau$ est formé des classes à gauche
$I_\tau(\sigma 
H)=I_\tau(\sigma)I_\tau(H)$ o\`u $\sigma H$ parcourt $L/H$ (i.e. $I_\tau$
induit une bijection de $L/H$ sur $L/H^\tau$). De même,
pour $g \in G$ et $C \in L/H$,
nous avons $I_\tau(g C)=I_\tau(g)I_\tau(C)$. 
Donc, en ordonnant correctement les classes à gauche,
l'action de $G$ sur $L/H$ est identique \`a celle de $G^\tau$
sur 
$L/H^\tau$. Plus précisément, en notant $\Theta$ la représentation de $G$ dans
$L/H^\tau$, nous avons $\Theta=I_\tau \Psi I_\tau^{-1}$.
D'o\`u $Gr(G,H)=Gr(G^\tau,H^\tau)=Gr(G^\tau,H)$, d'apr\`es la première
partie de cette démonstration. Donc l'indépendance du choix de $G$ est démontrée.
\end{proof}

Soient $H_1,\ldots ,H_r$ des représentants des classes de conjugaisons de 
$L$. Les matrices

$$
\MFP=(P_L(H_i,H_j))_{1\leq i, j \leq r} \quad \text{et} \quad 
\MFG=(Gr_L(H_i,H_j))_{1\leq i, j \leq r}
$$

\vspace{0.1in}
sont respectivement appelées la {\it matrice des partitions relative \`a $L$}
et la {\it matrice des groupes relative \`a $L$}.

\begin{propo} \label{propo : partpart} {\ }\\
1. La partition $P_L(G,H)$ est de la forme $1^m \ldots $ avec $m \geq 1$ si et
seulement si $G$ est un sous-groupe d'un conjugué de $H$ dans $L$ ; \\
2. $P_L(G,I_n)=\Card(H)^c$ où $c$ est l'indice de $G$ dans $L$ ;\\
3. $P_L(G,H)=[L : H]$ ssi $L=GH$ (par ex., $G=L$) ;\\
4. $P_L(I_n,H)=1^e$.
\end{propo}
\begin{proof}
1. Si $G$ est un sous-groupe d'un conjugué $H^\prime$ de $H$ dans $L$ alors
   $\Psi(G).H^\prime=\{H^\prime\}$. D'où $P_L(G,H)=P_L(G,H^\prime)=1^m
   \ldots $ avec $m \geq 1$. Inversement, supposons qu'il existe $C=\tau H$ une
   classe de $L/H$ telle que 
 $\Psi(G).C=\{C\}$ (i.e. $P(G,H)$ possède au moins une part égale à 1) ;
pour tout $\sigma \in G$, nous avons $\sigma\tau H=\tau H$ ; ce qui est
 équivalent à $\sigma \in \tau H \tau^{-1}$.\\
2. Le groupe $I_n$ est d'indice $e=\Card(L)$ dans $L$. Pour toute
   classe $C=\tau I_n=\{\tau\}$ de $L/I_n$, le cardinal de l'orbite $\Psi(G).C$ est
   donc identique à
   celui de $G$. Comme la partition $P(G,I_n)$ est de poids $e$,
   le résultat est démontré.\\ 
3. C'est lorsqu'il n'y a qu'une seule orbite.\\
4. Car $\Psi(I_n).C =C$ pour toute classe $C$ de $L/H$.
\end{proof}

\begin{propo}
(\cite{ArnaudiesValibouze:97}) Les lignes de la matrice des
  partitions (et donc aussi des groupes) sont distinctes deux à deux.
\end{propo}
\begin{proof}
Montrons que les lignes qui correspondent à $G$ et $H$ sont
   distinctes si ces deux groupes ne sont pas $L$-conjugués. Nous
   ferons référence 
   aux assertions 1. et 2. de la proposition \ref{propo : partpart}.
Si $G$ n'est pas un sous-groupe d'un conjugué de $H$ alors $P(G,H)\neq
   P(H,H)$, d'après 1.. Si $G$ est un
   sous-groupe propre d'un conjugué de $H$ alors, d'après 2., $P(G,I_n)\neq
   P(H,I_n)$. Si les lignes correspondant à $G$ et $H$  sont identiques
alors $P(G,H)=P(H,H)$ et $P(G,I_n)=P(H,I_n)$ et, par conséquent, $G$
   est un conjugué de $H$ dans $L$. 
\end{proof}

\begin{ex} \label{ex : 6.9} Pour $L=M_5$ et en reprenant les notations de
l'exemple \ref{ex : 6.3},
nous obtenons~:

$$
\MFG= \left( \begin{array}{llllll}
S_1^{20} & S_1^{10} & S_1^5 & S_1^4 & S_1^2 & S_1  \\
S_2^{10} & S_1^2,S_2^4 & S_1,S_2^2 & S_2^2 & S_1^2 & S_1  \\
{\mathcal H}_3^5 & S_2,{\mathcal H}_3^2 & S_1,{\mathcal H}_3 & {\mathcal H}_3 & S_2 & S_1  \\
H_4^4 & H_4^2 & H_4 & S_1^4 & S_1^2 & S_1  \\
{\mathcal H}_5^2 & H_5^2 & H_5 & S_2^2 & S_1^2 & S_1  \\
H_6^{(20)} & H_6^{(10)} & H_6 & {\mathcal H}_3 & S_2 & S_1 
\end{array} \right )
$$
où \\
- ${\mathcal H}_3$ est le sous-groupe cyclique de $S_4$ engendré par
$(1,4)(2,3)$ et $(1,2,4,3)$ ; on a $S_1\times{\mathcal H}_3 =H_3$,\\
- ${\mathcal H}_5=\gid{( 1, 2, 4, 7,10)( 3, 6, 9, 8, 5),
  ( 1, 3)( 2, 5)( 4, 8)( 6,10)( 7, 9) }$ est la représentation régulière
symétrique de
  $H_5$ (dans $S_{10}$),\\
- $H_6^{(10)}=\gid{( 1, 2, 4, 7,10)( 3, 6, 5, 9, 8),
  ( 1, 3, 7, 6)( 2, 5, 4, 8)( 9,10)}$ est une représentation symétrique de
  $H_6$ dans $S_{10}$,\\
- $H_6^{(20)}=\gid{( 1, 2, 4, 7,10)( 3, 6, 9, 8, 5),
  ( 1, 3)( 2, 5)( 4, 8)( 6,10)( 7, 9)}$ est la représentation symétrique
  régulière de $H_6$ dans $S_{20}$.
\end{ex}

{\bf Note} Tandis qu'E.H Berwick et H.O. Foulkes construisent des
sous-matrices de $\MFP$ pour $L=S_n$ ($n=5,6$ et 7), $G$ et $H$ parcourant les
sous-groupes transitifs de $S_n$, G. Butler et J. McKay prennent pour
$L$ les groupes symétriques jusqu'au degré 11, pour $G$ les
sous-groupes transitifs de $S_n$ et pour $H$ des groupes de la forme
$U \times S_m$ où $U$ est ou bien le groupe identité ou bien le groupe
symétrique de degré $n-m$. Dans ce qui est proposé ici, tous les
groupes sont considérés et nous ne calculons pas seulement $\MFP$ mais
aussi $\MFG$. Néanmoins, tous ces travaux s'inscrivent dans la même
démarche.

\section{Résolvante Générique}

Pour $P \in k[x_1,\ldots ,x_n]$, l'orbite de $P$ sous l'action de $L$
est l'ensemble $L.P$ suivant~:
$$ L.P =\{\sigma.P \; \mid \; \sigma \in L\} \quad .
$$

La {\it résolvante $L$-relative
  générique par $P$} est le
polynôme~:
$$
\calr(\underline{x},x) = \prod_{Q\in L.P}(x - Q) \quad .
$$

Supposons que $H$ soit le sous-groupe de $L$ stabilisant
$P$ dans $L$~: 
$$H =\{\sigma \in L \; \mid \; \sigma.P=P\} \quad (i.e. \; \; \text{Stab}_L(P)=H)
\quad .$$
Le polynôme $P$ est alors appelé un {\it $H$-invariant $L$-primitif}. La
proposition \ref{propo : 6.13} justifiera cette terminologie.

La proposition suivante est considérée comme classique.

\begin{propo} \label{propo : 6.11} {\ }\\
1.  L'orbite $L.P$ est constituée
  des 
$e$ polynômes distincts $\sigma.P$ où $\overline{\sigma}$ parcourt $L/H$. \\
2. Pour tout $\sigma \in L$, le polynôme $\sigma.P$ est un $H^\sigma$-invariant
$L$-primitif.
\end{propo} 

\begin{proof} Soient $\tau,\sigma \in L$. \\
1. Nous avons $\tau.P = \sigma.P$ si et seulement
si 
$\sigma^{-1}\tau.P=P$ ; ce qui est équivalent \`a $\tau \in \sigma H$.\\
2. De la même manière, nous avons $\tau.(\sigma.P)= \sigma.P$ si et seulement
si $\tau \in \sigma
H\sigma^{-1}$. 
\end{proof}

L'application 
$$ \begin{array}{lccc}
h_1 :& L/H &\longrightarrow & L.P \\
&\sigma H & \mapsto & \sigma.P 
\end{array}
$$
est une bijection entre $L/H$ et l'ensemble des racines de $\calr$. La représentation naturelle
$\Psi_1 $ de $G$ dans $S_{L.P}$ est équivalente à sa
représentation $\Psi$ dans $S_{L/H}$ : 
$$\Psi=h_1^{-1}\Psi_1 h_1 \quad .$$
 En effet,
pour 
tout $\tau \in G$ et $C =\sigma H \in L/H$ si $\tau C = C^\prime
=\sigma^\prime H$ alors $\tau.(\sigma.P)=\sigma^\prime .P$.

\begin{conv} Nous ordonnons l'orbite $L.P$ de telle sorte que le $i$-ième
élément noté $P_i$ soit l'image
par $h_1$ de la $i$-ième orbite $C_i$ de $L/H$.
\end{conv}

La représentation symétrique de $G$ dans
$S_e$ induite par $\Psi_1$ est identique à sa représentation $\Psym$ induite
par $\Psi$ (voir Paragraphe \ref{para : 1}) ~:
$$\begin{array}{cccl}
\Psym : &G  & \longrightarrow & S_e\\
&\tau   & \mapsto   &  \sigma : \sigma(i)=j \text{ si } \tau C_i = C_j
\; (\text{ i.e. } \tau.P_i = P_j)
                       \quad .
\end{array}
$$

D'après la proposition \ref{propo : 6.11}, nous pouvons définir
l'orbite $(L/H).P$ et nous 
avons~: 
$$
\calr = \prod_{\overline{\sigma} \in L/H}(x-\sigma.P) =
\prod_{Q \in (L/H).P}(x-Q) \quad .
$$

\begin{rem} \label{rem : essential} Soit $G$ est un sous-groupe de $L$.
Par la théorie de Galois classique, nous constatons que
  l'action à gauche de $G$ sur les classes de $L/H$ fournit les degrés
  et groupes de Galois des facteurs de $\calr$ sur le corps
  $K(x_1,\ldots ,x_n)^G$. Il s'agit donc des listes $P_L(G,H)$ et
  Gr$_L(G,H)$.
\end{rem}

\begin{exs} \label{ex : 6.12} {\ }\\
{\bf 1. (suite)} Le Vandermond $P=\prod_{1\leq i<j\leq 4}(x_i-x_j)$ est un
$A_4$-invariant $S_4$-primitif. Comme $(3,4).P=-P$, nous avons
$$
\calr = x^2 - P^2=x^2 - \prod_{1\leq i\leq j\leq
  4}(x_i-x_j) = x^2 -\Delta
$$
où $\Delta$ est le discriminant du polynôme $(x-x_1)(x-x_2)(x-x_3)(x-x_4)$.\\

{\bf 2. (suite)} Les polynômes $P=x_1+x_2$ et $Q=x_1x_2$ sont des
$S_{2^2}$-invariants $S_4$-primitifs. En se basant sur l'orbite $\mathcal O$,
il vient
$$
(S_4/S_{2^2}).P = \{x_1+x_3,x_2+x_4,x_1+x_2,x_2+x_3,x_3+x_4,x_1+x_4\} \quad .
$$

{\bf 3. (suite)} Le polynôme  $P=x_4x_5+x_3x_4+x_2x_3+x_1x_5+x_1x_2$ est un
$D_5$-invariant $M_5$-primitif et
$$
\calr = (x-P)(x-(2,3,5,4).P) = (x-P)(x-(x_2x_4+x_5x_2+x_3x_5+x_1x_4+x_1x_3))
\quad .
$$
\end{exs}

\'Etudions la résolvante $\calr$ comme un polynôme de $k(x_1,\ldots ,x_n)[x]$.
Notons $K$ le corps 
$$k(x_1,\ldots ,x_n)^{S_n}=k(\sigma_1,\ldots ,\sigma_n) ,$$
où $$\sigma_1=\sum x_i,\ldots ,\sigma_i=\sum x_1x_2\cdots
x_i,\ldots ,\sigma_n=x_1x_2\cdots x_n$$
sont les fonctions symétriques élémentaires
de $x_1,x_2,\ldots ,x_n$.  

\begin{propo} \label{propo : 6.13}  
Le polynôme $P$ est un élément primitif du corps
$K(x_1,\ldots ,x_n)^H$ sur le corps ${\mathcal K}= K(x_1,\ldots ,x_n)^L$ et
la résolvante $\calr$ est son
polynôme minimal sur le corps $\mathcal K$.
\end{propo}

\begin{proof} 
Montrons d'abord que $\calr$ est le polynôme minimal de $P$.\\
Les coefficients de $\calr$ étant des fonctions symétriques 
des éléments de $L.P$, ils sont invariants par $L$ et appartiennent donc
au corps $\mathcal K$. Montrons que $\calr$ est irréductible sur
 $\mathcal K$. Supposons que $h$ soit le facteur unitaire de 
$\calr$ et irréductible sur
 le 
corps $\mathcal K$ tel que $h(P)=0$. 
 Donc
$h$ est invariant par l'action de $L$ sur $x_1,\ldots ,x_n$. D'où, 
pour tout $\sigma \in L$, $\sigma.P$ est une racine de $h$ ; ce qui impose à
$h$ d'être un
multiple de $\calr$. Par conséquent, $\calr=h$ et $\calr$ étant irréductible sur
$\mathcal K$, elle est le polynôme minimal de $P$.\\
Nous en déduisons la primitivité de $P$ :
d'après la proposition \ref{propo : 6.11}, le degré de la résolvante $\calr$ est $e$. Donc le
polynôme minimal de $P$ a pour degré le cardinal
de $L/H$, qui, d'après la correspondance galoisienne, est le degré   
de l'extension $K(x_1,\ldots
,x_n)^H$ du corps $\mathcal K$.
\end{proof}

\begin{theorem} \label{theorem : 6.14} Posons ${\mathcal K}=K(x_1,\ldots
,x_n)^L$.
Soit le sous-groupe normal de $L$ donné par 
$$J=\bigcap_{\sigma \in
  L}H^\sigma \quad .$$ Alors

1. le corps des racines $P_1,\ldots ,P_e$ 
de la résolvante générique $\calr$ est 
$$
{\mathcal K}(P_1,\ldots ,P_e)= {\mathcal K}(x_1,\ldots
,x_n)^J \quad ;
$$
2. toute représentation symétrique dans $S_e$ du groupe $L/J$
est une représentation symétrique et transitive dans $S_e$ du groupe de Galois de
$\calr$ sur ${\mathcal K}$.
\end{theorem}
\begin{proof}
D'après la proposition \ref{propo : 6.13}, nous avons
$$
{\mathcal K}(P_1,\ldots ,P_e)= \bigcup_{\sigma \in L}{\mathcal
  K}(\sigma.P)=\bigcup_{\sigma \in L}{\mathcal K}(x_1,\ldots 
,x_n)^{H^\sigma}= {\mathcal K}(x_1,\ldots
,x_n)^J \quad .
$$
car, pour tout $\sigma \in L$, le polynôme $\sigma.P$ est un $H^\sigma$-invariant
$L$-primitif (voir Proposition \ref{propo : 6.11}) et $\calr$ est
également la résolvante par $\sigma.P$. La 
représentation est transitive puisque
$\calr$ est irréductible sur $\mathcal K$.
\end{proof}

Lorsque $L=S_n$ et $G$ est un sous-groupe transitif de $S_n$,
le théorème \ref{theorem : 6.14} induit les résultats suivants~:

\begin{itemize}
\item si $H=S_n$ alors $P\in K$ est un polynôme symétrique en $x_1,\ldots
  ,x_n$, $e=1$, $J=S_n$ ; $S_1$ est le groupe de Galois sur $K$ de la résolvante $\calr=x-P$~;

\item si $H=A_n$ alors $e=2$, $J=A_n$ et $S_2$, la représentation
  symétrique dans dans $S_e$ du groupe $S_n/A_n$, est le
groupe de Galois sur $K$ de la résolvante $\calr$~; nous avons
$$K(P_1,P_2)=K(P_1)\quad ; $$
comme $A_n$-invariant $S_n$-primitif $P$, nous pouvons prendre le Vandermond 
$$\delta=\prod_{1\leq
  i<j\leq n}(x_i-x_j) \quad  \text{ ; \; d'où } \quad \calr=x^2-\delta^2$$
  où $\delta^2$ est 
le discriminant du polynôme générique $\prod_{i=1}^n(x-x_i)$ ;

\item si $n=4$ et $H=V_4=\gid{(1,4)(2,3),(1,2)(3,4)}$ alors $e=6$,
  $J=V_4$ et le groupe de Galois
sur $K$ de la résolvante $\calr$ de degré 6  est isomorphe au groupe $S_4/V_4$ 
d'ordre 6 ; une représentation symétrique dans $S_6$ de ce groupe est
  le groupe
  $\gid{(1,2)(3,5)(4,6),(1,3)(2,4)(5,6)}$ ; on peut prendre
  $P_1=(x_1-x_2)(x_3-x_4)$  et on a :
  $$K(P_1,P_2,\ldots ,P_6)=K(P_1)=K(x_1,x_2,x_3,x_4)^{V_4}\quad
  ;$$  

\item si $n=4$ et $H=D_4$ alors $e=3$, $J=V_4$ et 
  $$K(P_1,P_2,P_3)=K(x_1,x_2,x_3,x_4)^{V_4} =K(P_i,P_j)\quad $$ 
pour tout $i\neq j$ ;
le groupe de Galois
sur $K$ de la résolvante $\calr$ de degré 3 est isomorphe au groupe $S_4/V_4$ 
d'ordre 6 ; $S_3$ est une représentation symétrique de ce groupe ; on peut
prendre $P=x_1x_3+x_2x_4$ ;

\item dans tous les autres cas, $J$ est le groupe identité 
et $S_n$ est isomorphe au
  groupe de Galois sur $K$
  de la résolvante $\calr$ ; nous avons alors
  $$K(P_1,P_2,\ldots 
  ,P_e)=K(x_1,\ldots ,x_n)\quad .$$
\end{itemize}


\section{Spécialisation de la résolvante générique}

Rappelons que $f=(x-\alpha_1)\cdots (x-\alpha_n)$ est un polynôme de $k[x]$ et que
$\Salpha=(\alpha_1,\ldots ,\alpha_n)$. Nous conservons les notations du
paragraphe précédent. En par\-ti\-culier, $H$ est le sous-groupe de
$L$ stabilisant l'invariant $P$~:
$$
\text{Stab}_L(P) = H\quad .
$$
Nous supposons que $G$ est le groupe de Galois de $\Salpha$ sur $k$.

La {\it résolvante $L$-relative de $\Salpha$ par $P$}
est le polynôme d'une variable~:
$$
R(x) = \calr(\Salpha,x) =\prod_{\overline{\sigma} \in L/H}
(x-(\sigma.P)(\alpha_1,\ldots ,\alpha_n))
\quad .
$$
Ce polynôme est aussi appelé une {\it $H$-résolvante $L$-relative de
  $\Salpha$.} \\

{\bf Note} Lorsque $L=S_n$, la résolvante ne dépend pas de la numérotation des racines de
  $f$ et elle peut s'appeler la {\it résolvante
  (absolue) 
de $f$ par $P$} ou une {\it $H$-résolvante (absolue) de
  $f$}. J.L. Lagrange a introduit la résolvante absolue. Afin de
  déterminer le groupe de Galois par descente dans le graphe
  d'inclusions des sous-groupes de $S_n$, R.P. Stauduhar utilise les
  résolvantes relatives avec $G\subset L$ (voir \cite{Stauduhar:73}) .

En toute généralité, les coefficients
de la résolvante $R$ appartiennent au corps $k(\Salpha)^{G\cap L}$. Le polynôme
minimal de $P(\Salpha)$ sur ce corps est donc
un facteur de la résolvante $R$. La proposition suivante énonce
un cas d'égalité.

\begin{propo} Posons $\beta=P(\Salpha)$ et prenons pour $L$
  un sous-groupe de $G$. La résolvante $R$ est le
 polynôme minimal de $\beta$ sur le corps 
 $k(\Salpha)^{L}$ (i.e. elle est irréductible sur ce corps) 
si et seulement si $\beta$ est une racine simple de $R$. Dans ce cas,
 $\beta$ est un
 élément $k(\Salpha)^{L}$-primitif du corps $k(\Salpha)^{H}$.
\end{propo}
\begin{proof} 
C'est une reformulation de la proposition \ref{propo : minbeta} car
les deux assertions sont équivalentes à Stab$_L(\beta)=H$.
\end{proof}

\begin{rem} 
Dans tout ce qui suit, on peut remplacer le corps $k$ par
 toute extension de $k$ et $G$ par le groupe de Galois de $\Salpha$
 sur ce corps. Ceci vaut, en particulier, pour toute extension
 intermédiaire entre $k$ et $k(\Salpha)$. 
\end{rem}

{\bf Hypothèse} Forts de la remarque précédente, nous 
supposons désormais que $G$ est un
sous-groupe de $L$ (i.e. $k=k(\Salpha)^{G\cap L}$).

Par le Théorème de Galois,
la résolvante $R(x)$ est à coefficients dans $k$ puisque ceux de
$\calr(\underline{x},x)$  
sont invariants par $L$. En particulier, si $R$ est une résolvante absolue, ses
coefficients sont des fonctions symétriques des racines de $f$ ; il
existe de nombreuses méthodes pour les calculer (indépendemment
de $G$) ;
certaines sont évoquées dans les articles de la
bibliographie. Le lecteur y trouvera aussi des méthodes
pour calculer des résolvantes $L$-relatives.

\begin{theorem} (\cite{ArnaudiesValibouze:97})
\label{theo : 6.15}
Supposons que $f$ soit sans racine multiple et que
  le corps $k$ soit infini. Il existe une $H$-résolvante
$L$-relative de $\Salpha$ sur $k$ qui soit sans racine
  multiple. Le polynôme $H$-invariant $L$-primitif associé à cette
  résolvante est alors dit $L$-séparable pour $\Salpha$.
\end{theorem}
\begin{proof} Toute résolvante $L$-relative de $\Salpha$
par $P$ étant un facteur sur $k$ de la résolvante de $f$ par $P$, nous pouvons
supposer que $L=S_n$.\\
Tout d'abord, montrons le théorème pour le groupe $H=I_n$. Soient 
$t_1,\ldots ,t_n$ des indéterminées et le polynôme
$$
V_{\underline{t}}=t_1x_1+\ldots + t_nx_n \quad .
$$
Pour toute permutation $\sigma$ de $S_n$ distincte de l'identité, nous avons
$$
V_{\underline{t}}(\alpha_{\sigma(1)},\ldots ,\alpha_{\sigma(n)}) \neq
V_{\underline{t}}(\alpha_1,\ldots ,\alpha_n) \quad 
.
$$

Le corps $k$ étant infini, il existe des valeurs
$\tilde{t_1},\ldots ,\tilde{t_n}$ de $k$ telles que 
$$
V(\alpha_{\sigma(1)},\ldots
,\alpha_{\sigma(n)})\neq V(\alpha_1,\ldots
,\alpha_n) \quad .
$$
où $V=V_{\tilde{t_1},\ldots ,\tilde{t_n}}$ est un polynôme de $k[x_1,\ldots
,x_n]$ (il n'existe qu'un nombre
fini de valeurs pour lesquelles il y a égalité). La résolvante $R$ de $f$ par
$V$ est donc une $I_n$-résolvante de $f$ sans racine multiple.\\
Soient $\tau_1H,\ldots \tau_eH$, $\tau_1=id$,
les classes à gauche de $L$ modulo $H$. Pour $i\in \intervalle{1}{e}$, posons
$$
R_i = \prod_{\tau \in \tau_iH}(x - \tau.V)
\quad .
$$
Les polynômes $r_i = R_i(\alpha_1,\ldots ,\alpha_n)$ sont des facteurs de la
$I_n$-résolvante séparable $R$~: $R=r_1r_2\ldots r_e$. Donc si $i\neq 1$ alors 
$r_1(x)\neq r_i(x)$. Le polynôme $r_1-r_i$ ne pouvant posséder plus de racines
que son degré, il existe une infinité de $u \in k$ tels que
$r_1(u) \neq r_i(u)$. Pour $u \in k$ bien choisi, le polynôme 
$R_1(u)$ est un $H$-invariant $S_n$-primitif et le polynôme
$\prod_{i=1}^e(x-r_i(u))$, résolvante de $f$ par 
$R_1(u)$, est sans racine multiple.
\end{proof}

\begin{note} \label{note : prim} Dans la démonstration
  précédente, lorsque $k$ est infini, $v$ est l'élément
 primitif de $k_1=k(\Salpha)$ 
 sur $k$ dont nous avons supposé l'existence au paragraphe
 \ref{section : corgal}. Lorsque $k$ est fini, il suffit de prendre un
 générateur du groupe fini monogène $k_1^*$.
De plus, si $H$ est un sous-groupe de $G$ alors
$r_i$ est un élément
$k$-primitif du corps Inv$(H)=k(\Salpha)^H$ lorsque le corps $k$ est
 infini. 
De même, lorsque $k$ est
 fini, Inv$(H)^*$ est un groupe fini monogène. Ceci constitue une
 démonstration du 
 théorème de l'élément primitif.
\end{note}

{\bf Note} Il existe des résolvantes génériques qui restent
séparables quelques soient 
les valeurs distinctes en lesquelles elles sont spécialisées.
C'est le cas de la $A_n$-résolvante $x^2 -
\delta^2$ et de la $M_5$-résolvante dite de Cayley 
(voir \cite{Cayley:61}).

\begin{exs} \label{ex : 6.16}
Les polynômes proviennent de la base de données du logiciel Magma.

{\bf 1. (suite)} Le polynôme $f=x^4 - x^3 - 3x^2 + x + 1$ possède $D_4$ comme groupe
de 
Galois sur $\Q$ et $R= x^2 -725$ où $725$ est le discriminant de $f$.

{\bf 2. (suite)} Gardons $f=x^4 - x^3 - 3x^2 + x + 1$. La résolvante par $P$
est le polynôme
$$
 R_1=x^6  - 3 x^5  - 3 x ^4 + 11 x^3 - 2 x^2  - 4 x + 1 \quad 
$$
et celle par $Q$ est le polynôme
$$
R_2 = x^6  + 3 x^5  - 2 x^4  - 8 x^3  - 2 x^2  + 3 x + 1 \quad .
$$
{\bf 3. (suite)} Le polynôme 
$f=x^5 - x^4 - 4x^3 + 3x^2 + 3x - 1$ possède $C_5$ 
comme groupe de Galois sur $\Q$. 
Si $C_5$ est le groupe de Galois de
$\Salpha=(\alpha_1, \ldots ,\alpha_5)$ sur $\Q$,
l'idéal $\MFM$ des $\Salpha$-relations est engendré par les 5 polynômes
\begin{eqnarray*}
 x_1^5 - x_1^4 - 4x_1^3 + 3x_1^2 + 3x_1 - 1, x_2 - x_1^3 + 3x_1,
    x_3 + x_1^2 - 2,\\
    x_4 - x_1^4 + x_1^3 + 3x_1^2 - 2x_1 - 1,
    x_5 + x_1^4 - 4x_1^2 + 2 \quad .
\end{eqnarray*}
Ces polynômes résultent de la factorisation de $f$ sur $\Q(\alpha_1)$ et sont
ordonnés de telle sorte que $C_5$ soit le groupe de décomposition de
$\MFM$ (i.e. $C_5.\MFM =\MFM$). Les évaluations des coefficients de $\calr$
modulo $\MFM$ donnent~: 
$$
R=(x+2)^2 \quad .
$$
\end{exs}

\section{Groupe de Galois de la résolvante}

Le degré des résolvantes $\calr$ et $R$ est $e$, l'indice de $H$ dans
$L$. Choisissons un ordonnancement $\beta_1,\ldots ,\beta_e$ des racines de
$R$ (que nous préciserons ultérieurement) et posons
$\Sbeta=(\beta_1,\ldots ,\beta_e)$.
Nous pouvons définir une représentation par permutations~:
$$
\begin{array}{lccl}
\Theta : & G &\longrightarrow & S_{\{\beta_1,\ldots ,\beta_e\}} \\
         & g & \mapsto        & \Theta(g) : \Theta(g).\beta_i = \beta_i^g \quad
         . 
\end{array}
$$

En effet, pour $g \in G$, d'une part, l'action $\beta_i^g$
est bien définie et d'autre part $\beta_i^g$ est bien une racine de $R$
puisque c'est 
une racine du polynôme minimal de $\beta_i$ sur $k$ qui est un
facteur de $R$. 

\begin{note} \label{note: multiso2}  Le sous-groupe $\Theta(G)$ de $G$
  n'est pas 
  nécessairement simplement isomorphe à $G$. En effet, supposons
  que 
  $g,g^\prime \in G$ satisfont $\beta_i^g =\beta_i^{g^\prime}$ et que
  les $\beta_i$ soient distincts deux-à-deux. Alors $g^\prime \in g
  J$ où $J$ est le sous-groupe normal 
$$\cap_{\sigma \in L} H^\sigma$$ de $L$. 
Le groupe $G$ est $m$-isomorphe au groupe 
$\Theta(G)$ où $m$ est l'ordre
du groupe $N=J\cap G$. Le groupe $G/N$ est simplement isomorphe à 
$\Theta(G)$ (Faire le lien avec la note \ref{note : multiso}).
Dans la littérature ancienne, nous retrouvons cette remarque sous
  diverses formes (voir, par exemple, \cite{Pierpont:1899})
\end{note}

Si la résolvante $R$ n'a aucune racine double,
la représentation symétrique du groupe $\Theta(G)$~:
$$\begin{array}{ccl}
\Theta(G) & \longrightarrow & S_e\\
\tau   & \mapsto   &  \sigma : \sigma(i)=j \text{ si } \tau.\beta_i = \beta_j
\end{array}
$$
est bien définie ; nous définissons ainsi une représentation symétrique de $G$
dans $S_e$~: 
$$
\begin{array}{lccl}
\Qsym ~: & G &\longrightarrow & S_e \\
         & g & \mapsto        & \sigma_g : \sigma_g(i) = j \text{ si }
         \beta_i^g = \beta_j \quad .
          
\end{array}
$$

\begin{conv} La représentation symétrique $\Psym(G)$ de $G$
  dans 
$S_e$ est induite par un ordonnancement
des classes de $L/H$ (voir Convention \ref{conv : 6.4}). Nous décidons que si
$\beta_j=\sigma.P(\Salpha)$ 
alors la $j$-ième classe est $\sigma H$ et qu'ainsi
  $\beta_j=P_j(\alpha_1,\ldots ,\alpha_n)$, où $P_j$ est le $j$-ième polynôme
  de l'ordonnancement choisi pour l'orbite $L.P$.
\end{conv}

\begin{theorem} \label{theo : 6.17} Supposons le corps $k$ infini.
Si la résolvante $R=(x-\beta_1)\ldots (x-\beta_e)$
  est sans racine multiple alors
la représentation symétrique $\Qsym(G)$ de $G$ dans
$S_e$ est le groupe de Galois de $\Sbeta$ sur $k$.
\end{theorem}

\begin{proof}
Notons $G_\Sbeta$ le groupe de Galois de $\Sbeta$ sur $k$. Montrons tout
d'abord que $\Qsym(G) \subset G_\Sbeta$. Soient $y_1,\ldots ,y_e$ des
variables.
Soit $g \in G$ et $p(y_1,\ldots ,y_e)$ un polynôme $G_\Sbeta$-invariant
$S_n$-primitif tel que pour tout $\sigma \not \in G_\Sbeta$ 
$$p(\beta_1,\ldots
,\beta_e)\neq (\sigma.p)(\beta_1,\ldots ,\beta_e) \quad .$$
Comme $k$ est infini, un tel polynôme existe
(Voir Théorème \ref{theo : 6.15}). Nous avons $p(\beta_1,\ldots
,\beta_e) \in k$ car $p$ est invariant par le groupe de Galois de $\Sbeta$ sur
$k$. Posons $q=p-p(\beta_1,\ldots
,\beta_e)$ et $Q=q(P_1(x_1,\ldots ,x_n),\ldots
,P_e(x_1,\ldots ,x_n))$. Nous avons $0=q(\beta_1,\ldots
,\beta_e)=Q(\alpha_1,\ldots ,\alpha_n)$. Par définition du groupe de Galois de
$\Salpha$ sur $k$, nous avons $(g.Q)(\alpha_1,\ldots ,\alpha_n)=0$. Donc
\begin{eqnarray*}
0=(g.Q)(\alpha_1,\ldots ,\alpha_n)& = & Q(\alpha_{\sigma(1)},\ldots
,\alpha_{\sigma(n)})\\
&= & q(g.P_1(\alpha_1,\ldots ,\alpha_n),\ldots ,g.P_e(\alpha_1,\ldots
,\alpha_n))\\
&=& q(\beta_1^g,\ldots ,\beta_e^g)\\
&=& (\Qsym(g).q)(\beta_1,\ldots ,\beta_e) \quad .
\end{eqnarray*}
Seules les permutations de $G_\Sbeta$ dans $S_e$ envoient la $\Sbeta$-relation $q$
sur une autre 
$\Sbeta$-relation. Donc $\Qsym(G)$ est un sous-groupe de $G_\Sbeta$.

Pour montrer l'inclusion inverse, choisissons un polynôme $p(y_1,\ldots ,y_e)$
qui soit un $I_e$-invariant $S_e$-primitif et tel que pour toute permutation
$\sigma \in S_n$ distincte de l'identité $\sigma.p(\beta_1,\ldots
,\beta_e)\neq p(\beta_1,\ldots ,\beta_e)$. Comme $k$ est infini, un tel polynôme existe (Voir
Théorème \ref{theo : 6.15}). Comme $p(\Sbeta)$ est un élément
$k$-primitif de $k(\Sbeta)$, son polynôme minimal sur $k$ est~:
$$
M=\prod_{\sigma \in G_\Sbeta}(x - (\sigma.p)(\beta_1,\ldots ,\beta_e)) \quad .
$$
C'est, en fait, la résolvante de Galois de la résolvante $R$. Nous avons
donc également, par la théorie de Galois et en posant $\gamma
=p(\beta_1,\ldots ,\beta_e) \in 
k(\alpha_1,\ldots ,\alpha_n)$,~:
$$
M=\prod_{\theta \in \{\gamma^g \mid g \in G\}}(x-\theta) \quad .
$$
En procédant comme dans la première partie de cette démonstration, nous
obtenons que 
pour tout $\sigma \in G_\Sbeta$ il existe $g \in G$ tel que 
$$
(\sigma.p)(\beta_1,\ldots ,\beta_e)=\gamma^g=(\Qsym(g).p)(\beta_1,\ldots
,\beta_e) \quad .
$$
Comme $\sigma^{-1} \in G_\Sbeta$, nous pouvons écrire~:
$$
(\sigma^{-1}\Qsym(g).p)(\beta_1,\ldots
,\beta_e)=p(\beta_1,\ldots
,\beta_e) \quad ;
$$
ce qui, par le choix de $p$, impose que $\Qsym(g) =\sigma$. D'où $G_\Sbeta
\subset \Qsym(G)$ et le théorème est démontré.
\end{proof}

{\bf Note} Dans la première partie de la 
démonstration précédente, il est possible d'utiliser une
variante montrant que  
le groupe $\Qsym(G)$ envoie toute $\Sbeta$-relation sur une autre
$\Sbeta$-relation. Nous avons choisi de prendre une relation particulière
possédant les propriétés nécessaires et suffisantes à la description de l'idéal
des $\Sbeta$-relations engendré par les modules de Cauchy
de la résolvante (i.e. des relations symétriques) et par la
$\Sbeta$-relation $q$ (voir Note \ref{note : geneideal}).


{\bf Note} Tout ceci est cohérent car, étant donné $\gamma \in k(\beta_1,\ldots
,\beta_e)\subset k(\alpha_1,\ldots ,\alpha_n)$, les permutations $\sigma \in G_\beta$ et $g \in
G$  de la démonstration précédente satisfont~:
$$
\gamma^\sigma = \gamma^g
$$
si la représentation du groupe $Aut_k(k(\beta_1,\ldots
,\beta_e))$ dans $S_e$ est celle associée à $\Sbeta$.

Si les racines de $R$ sont distinctes deux-à-deux (i.e. $R$ est séparable),
l'application 
$$ \begin{array}{lccl}
h_2 :& L.P &\longrightarrow &  \{\beta_1,\ldots ,\beta_e\} \\
&Q & \mapsto & Q(\alpha_1,\ldots ,\alpha_n) \quad .
\end{array}
$$
est une bijection entre les racines de $\calr$ et celles de $R$,
l'application $h=h_2oh_1$ est alors une bijection de $L/H$ dans 
$\{\beta_1,\ldots ,\beta_e\}$ et
$$
\Theta = h \Psi h^{-1} \quad .
$$
Les représentation $\Theta$ et $\Psi$ étant ainsi équivalentes, les
représentations 
$\Qsym$ et $\Psym$ de $G$ dans $S_e$ sont identiques. On en déduit
le théorème suivant qui pré-détermine le groupe de Galois
de $R$ uniquement à partir de $G,H$
et $L$.

\begin{theorem}\label{theo : 6.18} 
Si la résolvante 
$R$ est sans racine multiple alors
la représentation symétrique $\Psym(G)$ de $G$ dans
$S_e$ est le groupe de Galois de $\Sbeta$ sur $k$ ; i.e. c'est le 
groupe de Galois de $R$ sur $k$.
\end{theorem}

Dans le cas où le groupe de Galois de $f$ est inconnu et celui le $R$
l'est partiellement, ce corollaire permet de savoir si $G$
n'est pas identique à certains sous-groupes
de $S_n$. En effet, l'ensemble des groupes $\Psym(G^\prime)$ où $G^\prime$
parcourt $S_n$ est pré-calculable (voir Paragraphe 1).
Une information partielle du groupe de Galois de $R$ est, par
exemple, celle des groupes de Galois de ses facteurs sur $k$. C'est à cette
information qu'est consacré le paragraphe suivant.

\section{Groupes de Galois des facteurs de $R$ et détermination de $G$}

Les orbites de $\Psi(G)$ sont en bijection avec celles de $\Theta(G)$. Le
groupe $G$ étant le groupe de Galois de $\Salpha$ sur $k$, l'ensemble des
orbites de 
$\Theta(G)$ est en bijection avec l'ensemble des facteurs irréductibles (pas
nécessairement simples) sur
$k$ de la résolvante $R$ : si $R(\beta)=0$ alors
$$\Theta(G). \beta \mapsto \prod_{\gamma \in
  \Theta(G). \beta}(x-\gamma) = {\text Min}_{\beta,k} \quad .$$


Soit $\beta=(\sigma.P)(\alpha_1,\ldots ,\alpha_n)$ une
  racine de la résolvante $R$, $C=\sigma H$ et $g_1=id,\ldots ,g_c$ des
  permutations de $G$ telles que~:\\
- $\Psi(G).C=\{C=g_1C,\ldots
  ,g_cC\}$\\
- si $C$ est la $j$-ième classe de $L/H$ (voir Convention \ref{conv : 6.4}) alors $g_iC$
  est la $j+i-1$-ième classe ; c'est-à-dire que
  $\beta^{g_i}=P_{j+i-1}(\alpha_1,\ldots ,\alpha_n)$.\\

Dans ce paragraphe et le suivant, nous considérerons 
le polynôme 
$$
F = \prod_{i=1}^{c}(x- \beta^{g_i})
$$
de $k[x]$. Ce polynôme est une puissance du polynôme minimal de 
$\beta$ sur $k$. S'il est sans racine multiple alors il est irréductible sur
$k$. 

\begin{theorem}\label{theo : 6.19} {\it  
Supposons que $\Psi(G).C$ soit la $s$-i\`eme orbite  de 
  $\Psi(G)$.\\
Si le polynôme $F$ est sans racine multiple alors~:\\
- le groupe de
  Galois 
de $(\beta,\beta^{g_2},\ldots ,\beta^{g_c})$ sur $k$ est 
$\Psi(G)_s$, le $s$-ième
élément de la suite $Gr_L(G,H)$, et \\
- le degré $c$ de $F$ est la
$s$-ième part de la partition 
$P_L(G,H)$.\\
 Le groupe $\Psi(G)_s$ est donc une représentation
symétrique dans $S_c$ du groupe de Galois de $F$ sur $k$.}
\end{theorem}
\begin{proof} 
Nous avons $\beta^{g_i}=(g_i\sigma.P)(\alpha_1,\ldots
,\alpha_n)$ pour $i=1,\ldots , c$. Si $F$ est sans racine multiple,
l'ensemble des racines de $F$ est en bijection avec l'orbite $\Psi(G).C$. La
démonstration se termine avec la définition de $Gr_L(G,H)$ (voir
Notation-Définition \ref{nota : 6.6}).
\end{proof}

\begin{rem}\label{rem : 6.20}
Soit $F$ un facteur irréductible simple sur $k$
de degré $c$ de la résolvante
$R$. Alors, en choisissant $\beta$ tel que $F(\beta)=0$, les conditions du
théorème sont satisfaites et  
le groupe de Galois de $F$ sur $k$ est l'un des groupes de degré $c$ de
la suite $Gr_L(G,H)$ (à un isomorphisme près).
\end{rem}

{\bf Note} Il est intéressant de constater que si la résolvante est
sans racine double, elle est irréductible si et seulement si il n'y a
qu'une seule orbite pour $\Psi(G)$. C'est-à-dire lorsque $L=GH$ (voir
3. Proposition \ref{propo : partpart}). Il ne faut pas en être
étonné. Lorsqu'on considère l'idéal de Galois défini par $\Salpha$ et
le groupe $H$ alors le plus grand ensemble de permutations définissant
aussi cet idéal est $GH$ (qui n'est pas nécessairement un groupe).

Le corollaire suivant est utilisé par R.P. Stauduhar dans sa
descente des sous-groupes. Il peut aussi être déduit du théorème
\ref{theo : 6.24}. 

\begin{corollaire} \label{coro : 6.25} {\it Si $\beta=(\sigma.P)(\Salpha)$ est
  une racine simple sur $k$ de la
  résolvante $R$ alors $G$ est un sous-groupe du conjugué $H^\sigma$ de $H$
  dans $L$.}
\end{corollaire}

\begin{rem}\label{rem : 6.26}
Si $G$ est le groupe de Galois de $\Salpha$ sur $k$ alors $G^{\sigma^{-1}}$
est celui de $\sigma.\Salpha$ sur $k$. Pour $\sigma \in L$, la résolvante
$L$-relative de $\Salpha$ par $P$ et celle de  $\sigma.\Salpha$ sont
identiques. 
Donc, si $G \subset H^\sigma$ alors $G^{\sigma^{-1}} \subset H$. Il suffit
d'échanger $\Salpha$ et $\sigma.\Salpha$ pour que dans le corollaire
précédent le groupe de Galois soit inclus dans le groupe $H$.
Lorsque le groupe $H$ est distingué dans $L$ alors $G$ est un
sous-groupe de $H$. Nous retrouvons ainsi la propriété bien connue que
le groupe de Galois est pair si son discriminant est un carré dans $k$.
\end{rem}

\begin{corollaire}\label{coro : 6.22}
{\it 
Si la résolvante $R$ est séparable alors la suite
$Gr_L(G,H)$ est à une permutation près la liste des groupes de Galois sur $k$
des facteurs irréductibles de $R$ sur $k$ et $P_L(G,H)$ est celle de leurs
degrés.}
\end{corollaire}

\begin{theorem}\label{theo : 6.23}(\cite{ArnaudiesValibouze:97}) {\it Supposons que le corps $k$ soit infini.
 Il est toujours possible de déterminer le groupe de
Galois $G$ avec des résolvantes.}
\end{theorem}
\begin{proof}
Car les lignes de la matrice $\MFP$ sont distinctes deux-à-deux et qu'il
existe toujours 
des résolvantes séparables.
\end{proof}

Examinons le cas des racines multiples..
\begin{theorem} {\ } \label{theo : multi} Nous distinguons 2 cas de multiplicité :\\
i) Si $\beta$ est de multiplicité exactement $m$ dans $F$ alors $m$
divise $c$ et
  $$F=F_0^m$$
 où $F_0$ est irréductible sur $k$.\\
ii) Si $\beta$ est aussi une racine de $R/F$ alors $F^2$ divise $R$ ; plus
 précisément, $\beta$ est une racine du facteur $F$ de $R/F$
 associé à une orbite de $\Psi(G)$ distincte de $\Psi(G).C$ mais de
 même cardinalité.
\end{theorem}
\begin{rem}
Dans le cas ii), le théorème \ref{theo : 6.19} restant valide, ce
cas ne peut se produire que si le groupe $\Psi(G)_s$ apparaît deux
fois dans $Gr_L(G,H)$. 
\end{rem}
\begin{proof}
Soient $P_1$ et $P_2$ deux racines distinctes de la résolvante générique
$\mathcal R$. Si $P_1(\Salpha)=P_2(\Salpha)$ alors pour tout $g\in G$
$g.P_1$ et $g.P_2$ sont deux racines distinctes de $\mathcal R$ telles
que $g.P_1(\Salpha)=g.P_2(\Salpha)$.
Supposons que $\sigma.P=P_1$ (i.e. $\beta=P_1(\Salpha)$).\\
Montrons i). Si $P_1$ et $P_2$ sont dans la même $G$-orbite
(i.e. $G.P_1=G.P_2$) alors $P_1(\Salpha)$ et $P_2(\Salpha)$ sont deux
racines de $F$ de même que $g.P_1(\Salpha)$ et $g.P_2(\Salpha)$. Donc
si $\beta=P_1(\Salpha)$ est de multiplicité $m$ dans $F$ 
alors toute autre racine de $F$ (i.e. $g.P_1(\Salpha)$, avec $g\in
G$) est aussi de multiplicité $m$.\\
Montrons ii). Supposons que $\beta$ soit une racine commune à $F$ et à
$R/F$. On a 
$P_1(\Salpha)=P_2(\Salpha)$ avec $P_1$ et $P_2$ dans deux
$G$-orbites distinctes. Donc toutes les valeurs des deux $G$-orbites
s'identifient deux-à-deux. D'où le résultat.
\end{proof}

\begin{corollaire}{\ }\\
i) Si le degré de $F$ est un nombre premier alors $F$ est soit
irréductible sur $k$ soit une puissance d'un facteur linéaire sur $k$.\\
ii) Si $GL=L$ et que $e=[L :H]$ est premier alors la résolvante $R$ 
est soit irréductible sur $k$, soit une puissance une puissance d'un
facteur linéaire sur $k$.
\end{corollaire}

\begin{exs}\label{ex : 6.21}
Nous supposons que $f$ n'a que des racines simples.\\
{\bf 1. (suite)} La résolvante $x^2-\Delta$ par $A_4$ est nécessairement sans
racine 
multiple car le discriminant $\Delta$ de $f$ est non nul. 
Le discriminant 725 de $f$ se factorise en $5^2.29$ qui n'est
pas un carré dans $\Q$. La résolvante $R$ est irréductible sur $\Q$. Son
groupe de Galois est nécessairement $S_2$.
Nous avions $Gr_{S_4}(D_4,A_4)=S_2$ qui annon{\c c}ait ce résultat.\\

{\bf 2. (suite)} La factorisation de la résolvante par $P$ est
$$
R_1= (x^2  - x - 1) (x^4  - 2 x^3 - 4 x^2 + 5 x - 1)
$$
et celle par $Q$ est~:
$$
R_2= (x + 1)^2  (x^4  + x^3  - 5 x^2  + x + 1) \quad .
$$
D'après l'exemple \ref{ex : 6.7}, $Gr_{S_4}(D_4,S_2\times S_2)=S_2, D_4$. Donc $D_4$ est
le groupe de Galois sur $\Q$ de chacun des facteurs de degré 4
de $R_1$ et $R_2$. Le polynôme $x^2  - x - 1$ étant simple, son groupe de
Galois $S_2$ était également prévisible. Le facteur $(x+1)^2$ de $R_2$ 
provient d'une orbite de cardinal 2. C'est le cas i) du théorème
\ref{theo : multi}. 

{\bf 3. (suite)} La résolvante $R=(x+2)^2$ possède une racine double.
Mais nous savons
que chaque facteur correspond à chacune des orbites 
$\{D_5\}$ et $\{(2,3,5,4)D_5\}$ car le groupe de Galois est
$C_5$. C'est le cas ii) du théorème
\ref{theo : multi}. 
\end{exs}

\section{Corps des racines de la résolvante $R$}
\label{para : corpsracines}
\'Etudions les racines de la résolvante. Nous allons constater que de
spécialiser $x_i$ en $\alpha_i$, pour $i=1,\ldots ,n$,
revient à intersecter les groupes avec le groupe de Galois $G$ de $f$ sur $k$. 
Nous savons déjà que le corps $k(x_1,\ldots ,x_n)^{S_n}$ se spécialise en le
corps 
$k=k(\Salpha)^{S_n\cap G}$ car les fonctions symétriques des racines de $f$ 
appartiennent à $k$. Plus généralement, comme $G \subset L$, le corps
${\mathcal K}=K(x_1,\ldots ,x_n)^L$ se spécialise en
$k=k(\Salpha)^G=k(\Salpha)^{L\cap G}$.

\begin{theorem}\label{theo : 6.24}
{\it Si le polynôme
 $F$ est sans racine multiple alors sa racine $\beta=(\sigma.P)(\alpha_1,\ldots
 ,\alpha_n)$ est un élément $k$-primitif du  
 corps $k(\Salpha)^{G\cap H^\sigma}$ et $F$ est son  polynôme minimal sur
 $k$. Par conséquent, le corps $k(\Salpha)^{G \cap H}$ s'identifie à
l'ensemble des $P(\Salpha)$ où $P$ parcourt
  $K(x_1,\ldots ,x_n)^H$.}
\end{theorem}

\begin{proof}
Soit $g \in G$. Les égalités $\beta^\sigma = \beta$ et
$g\sigma.P=\sigma.P$ sont équivalentes puisque $F$ est sans racine multiple et 
que ses racines
sont les spécialisations en $\Salpha$ de la $G$-orbite de $\sigma.P$. Comme
$\sigma.P$ est un $H^\sigma$-invariant $L$-primitif et que $G$ est un
sous-groupe de $L$, $\beta^g = \beta$ est donc équivalent à $g \in
H^\sigma$. Soit $U$ le sous-groupe de $G$ tel que $k(\beta)=K(\Salpha)^U$.
Si $K(\Salpha)^U$ était strictement inclus dans le corps
$k(\Salpha)^{G\cap H^\sigma}$ alors il existerait $\tau \in U$ tel que
$\tau 
\not \in H^\sigma$ et
$\beta^\tau = \beta$ ; ce qui est impossible. D'où $U=H^\sigma$.
\end{proof}

{\bf Note} En rapprochant les théorèmes \ref{theo : 6.15} et
\ref{theo : 6.24}, on trouve une méthode 
pour calculer un élément $k$-primitif de tout sous-corps du 
corps des racines de $f$.

\begin{corollaire}
Si $G \cap H^\sigma$ est le groupe identité et que $F$ est sans
 racine multiple alors
 $\beta$ est un élément $k$-primitif du corps $k(\Salpha)$ et
 $F$ est son polynôme minimal sur $k$.
\end{corollaire}

\begin{corollaire}\label{coro : 6.27}
{\it Si le polynôme
 $F$ est sans racine multiple alors son groupe de Galois sur $k$ est
isomorphe au groupe $G/G\cap M$ où $$
M=\bigcap_{g \in G}H^{g\sigma} \quad .$$
 En
 particulier, si $G\cap M$ est le groupe identité alors $G$ est isomorphe au
 groupe de 
 Galois de $F$ sur $k$ et le corps des racines de $F$ est identique à celui de
 $f$.}
\end{corollaire}
\begin{proof} L'ensemble des racines de $F$ est formé
 des $\beta^g$ où $g$ parcourt $G$. Nous avons donc
\begin{eqnarray*}
k(\beta^{g_1},\beta^{g_2},\ldots ,\beta^{g_c})&=&\bigcup_{g \in G}k(\beta^g)\\
                                     &=& \bigcup_{g \in G}k(\Salpha)
                                                  ^{G\cap H^{g\sigma}}
\end{eqnarray*}
car les racines de F sont distinctes (voir Théorème 
\ref{theo : 6.24}). Donc le corps des 
racines de $F$ est 
identique au corps  $k(\Salpha)^{G \cap M}$. Par conséquent, le sous-groupe
normal  $G \cap M$ de $G$ est le groupe de
Galois de $k(\Salpha)$ sur le corps des racines de $F$.
\end{proof}

De la même manière, nous obtenons le corollaire suivant qu'il faut
rapprocher du théorème \ref{theorem : 6.14} et
des notes \ref{note : multiso} et \ref{note: multiso2}. Nous le
trouvons dans \\

\begin{corollaire}\label{coro : 6.28}
{\it Si la
 résolvante $R$ est sans racine multiple alors son groupe de Galois sur $k$ est
isomorphe au groupe $G/ G\cap J$ où 
$$J=\bigcap_{\sigma \in L}H^\sigma
 \quad .$$ En
 particulier, si $G \cap J$ est le groupe identité alors $G$ est le groupe de
 Galois de $R$ sur $k$ et le corps des racines de $R$ est identique à celui de
 $f$ :
$$
k(\Salpha) = k(\Sbeta) \quad 
$$
et le groupe de Galois de $f$ sur $k$ est isomorphe à celui de $R$ sur $k$.}
\end{corollaire}

Dans le corollaire suivant, nous excluons les cas dans lesquels $N$ n'est pas
le groupe identité (voir les exemples du paragraphe \ref{para : connu}).

\begin{corollaire}\label{coro : 6.29}
 (\cite{ArnaudiesValibouze:97}) {\it Supposons que $L=S_n$. Supposons
que $H \not\in \{S_n,A_n\}$ et que, de plus, si $n=4$ alors 
$H \not \in \{D_4,V_4\}$. Alors $k(\Salpha) = k(\Sbeta)$.}
\end{corollaire}

\section{Exemples de résolvantes connues}
\label{para : connu}
Ici, nous retrouvons des résultats classiques.

{\bf 1.} Supposons que le polynôme $f$ est irréductible et
que son degré $n$ est $4$.
Soit $P=x_1x_3 + x_2x_4$ un $D_4$-invariant $S_4$-primitif.
La résolvante $R$ de degré $e=3$
est connue sous le nom de {\it résolvante cubique}. Supposons que $R$ n'ait pas
de racine double. Nous avons
$$J=V_4 \quad .$$ Il y a cinq possibilités pour le groupe de
Galois $G$~:\\ 
- $G=S_4$ ;  $G \cap J = V_4$ ;\\
$S_4/V_4$ est d'ordre 6 ; donc $R$ est
irréductible sur $k$ de groupe de Galois $S_3$ ;\\ 
- $G=A_4$ ; $G \cap J =V_4$ ;\\
$A_4/V_4$ est d'ordre 3 ; donc $R$ 
est irréductible sur $k$ de groupe de Galois $A_3$ ; \\
- $G=V_4$ ; $G \cap J = V_4$ ;\\
$V_4/V_4$ est le groupe identité ; donc $R$ se
factorise en 3 facteurs linéaires sur $k$ ;\\
- $G=D_4$ ; $G\cap J =V_4$ et $G\cap H=D_4$ ;\\
$D_4/V_4$ est d'ordre 2 ; donc le groupe
de Galois de $R$ est $S_2$ ; comme $G/G\cap H$ est le groupe identité, la
racine $P(\alpha_1,\alpha_2,\alpha_3,\alpha_4)$ de $R$ appartient à $k$ ; $R$
se factorise sur $k$ en un facteur linéaire et un de degré 2 ;\\
- $G=C_4$ ; $G \cap J =\gid{(1,3)(2,4)}$ ;\\
$C_4/(G \cap J)$ est
d'ordre 2 ;  donc $R$ se factorise sur $k$ en un facteur de degré 2 et un
facteur linéaire de racine $P(\alpha_1,\alpha_2,\alpha_3,\alpha_4)$.\\

{\bf 2.} Soit $P \in k[x_1]$. La résolvante de $f$ par $P$ est appelée une {\it
  résolvante de Tchirnhaus}. C'est le polynôme de degré $e=n$~:
$$
R = \prod_{i=1}^n(x-P(\alpha_i)) \quad .
$$
Cette résolvante est sans racine multiple si $P(\alpha_i)\neq P(\alpha_j)$
pour $i\neq j$.
Le groupe $H$ est le groupe $S_1\times S_{n-1}$. Le groupe $J=\cap_{\sigma \in
  S_n}H^\sigma$ est le groupe identité. Si la résolvante est sans racine
multiple alors son groupe de Galois est 
identique à celui de $f$. En particulier, si $P=x_1$ alors $R=f$ et 
$Gr_L(G,H)$ est la liste des groupes de Galois des facteurs
irréductibles de $f$ sur $k$.

{\bf 3.} Soit $V=t_1x_1+\cdots +t_nx_n$, $t_i \in k$ distincts
deux-à-deux, un $I_n$-invariant 
$S_n$-primitif. Ici 
$H=I_n$ et $G \cap H^\sigma =I_n$ pour tout $\sigma \in S_n$.
La résolvante de $f$ par $P$
est un polynôme de degré $n!$ appelé {\it résolvante de Galois de
  $f$}. Supposons que $F$ soit un facteur 
simple de cette résolvante et que $\beta$ en soit un racine. Alors,
d'après le théorème \ref{theo : 6.24}, $\beta$ est un élément
$k$-primitif du corps  
$k(\alpha_1,\ldots ,\alpha_n)$ des racines de $f$. Le polynôme $F$ a pour
degré l'ordre du groupe de Galois de $f$ sur $k$ et il s'exprime ainsi~:
$$
F =\prod_{g \in G}(x-\beta^g) \quad .
$$
Le groupe de
Galois de $F$ sur $k$ est isomorphe à $G$, le groupe de Galois de $\Salpha$
sur $k$. Dans la littérature, la résolvante de Galois désigne parfois toute
résolvante $G$-relative
de $\Salpha$ par $V$ qui ne possède pas de racine multiple; c'est-à-dire le
polynôme $F$. 

{\bf 4.} 
Si le polynôme $f$ est unitaire sans racine multiple, une $A_n$-résolvante
  de $f$  
est le polynôme séparable $x^2-\Delta(f)$, où $\Delta(f)$ est le discriminant
  de 
  $f$. Le groupe de Galois $G$ de $\Salpha$ sur $k$ est pair si
  et seulement si le discriminant de $f$ est un carré sur $k$ (voir
  Remarque \ref{rem : 6.26}).

{\bf 5.} Soit $H=M_5$, le groupe métacyclique de degré 5. La résolvante de $f$
par $P$ est un polynôme de degré 6. Si elle est sans racine multiple, le corps
des racines de cette résolvante est identique à celui de $f$. 
Supposons que $f$ soit irréductible sur $k$ et que la résolvante n'ait pas de
racine multiple.
Si $G=S_5$, la résolvante est irréductible. Si $G=A_5$ alors la
résolvante possède un facteur irréductible de degré $[A_5 : A_5\cap
M_5]=60/10=6$ ; donc 
elle est irréductible. Sinon, pour une numérotation adéquate des racines de
$f$, le groupe $G$ est l'un des groupes $M_5,D_5$ ou $C_5$ et la
résolvante possède un facteur linéaire sur $k$ (et un de degré 5, si on étudie
cas par cas le degré de l'extension que doit diviser l'ordre du groupe $G$). 
Pour que $G$ soit un groupe résoluble
et que $f$ soit résoluble par radicaux il faut et il suffit que $G$
soit un sous-groupe de $M_5$.
Le polynôme
$$
P=x_1x_2+x_2x_3+x_3x_4+x_4x_5+x_5x_1 - x_1x_3 +x_3x_5+x_5x_2+x_2x_4+x_4x_1
$$
est un $M_5$-invariant $S_5$-primitif. La résolvante $R$ de $f$ par $P$ est
connue sous le nom de {\it résolvante de Cayley} (pour son expression,
voir \cite{Cayley:61} et \cite{LITP9361}).
Cette
résolvante est sans racine multiple si $f$ l'est également. Elle
permet donc toujours de
tester si un polynôme irréductible de degré 5 est ou non résoluble par
radicaux. 


{\bf Commentaires.}

1. Dans les exemples précédents, des résultats 
du paragraphe \ref{para : corpsracines} sont appliqués pour déterminer
les groupes ou degrés des facteurs de $R$. La matrice des groupes
fournit les mêmes informations. 

2. Pour $G=S_4$ ou $G=A_4$, en déterminant le groupe de
Galois d'un polynôme de degré 3, facteur irréductible de $R$, nous
déterminons celui du polynôme $f$ 
de degré $n= 4$. Cette chute du degré est un des avantages que présente la
matrice des groupes 
face à celle des partitions. Cette situation n'est pas rare. Elle
devient très avantageuse lorsque le degré $n$ s'élève (voir
\cite{Valibouze:95} pour le degré 8). 

\section{Applications des résolvantes et des matrices de groupes}

En se restreignant aux résolvantes absolues, il est toujours possible
d'identifier le groupe de Galois d'un polynôme non nécessairement
irréductible. Les calculs des 
résolvantes sont 
réalisés avec des manipulations de fonctions symétriques. Le problème
est que les degrés des résolvantes absolues nécessaires à discriminer les
groupes s'élèvent rapidement en fonction du degré $n$ du polynôme.

Il faut alors pouvoir calculer des résolvantes
relatives. R.P. Stauduhar propose de le faire avec des méthodes
numériques. Les résolvantes elles-mêmes offrent un moyen de calculer des
résolvantes relatives (voir \cite{AubryValibouze:00} et
\cite{Valibouze:99}) ; une autre 
méthode est proposée dans \cite{LITP9361}. Les résolvantes relatives
étant des facteurs de résolvantes absolues, le problème de la
croissance des degrés est ainsi contrôlé. Cette méthodologie a pour
avantage de calculer simultanément l'idéal $\MFM$ (i.e. le corps des
racines).Toujours dans l'idée de calculer l'idéal $\MFM$, dans 
\cite{Sargov:2003.05} les auteurs travaillent sur les facteurs du
polynôme $f$ dans ses extensions 
$k(\alpha_1,\ldots,\alpha_i)$. Les matrices de groupes sont
encore utilisables dans ce cadre. En particulier, $f$ étant une
résolvante sa factorisation donne des informations sur le groupe de
Galois $G$. 

Les matrices de groupes sont utilisable en sens inverse pour calculer
un polynôme de degré $c$ dont le groupe de Galois $\mathcal G$ apparaît dans
$G_L(G,H)$ : supposons que $f$ est donné avec son groupe $G$ ; on calcule
une $H$-résolvante $L$-relative de $f$ dont le facteur associé à
l'orbite induisant $\mathcal G$ dans $G_L(G,H)$ est sans racine
multiple. Ce facteur est le polynôme cherché. Cette méthode proposée
dans \cite{Valibouze:95} a été
appliquée dans 
\cite{GilVal:96} avec $m=12$. Elle est aussi
utilisable pour calculer des polynômes dans des extensions de $k$.

Les matrices de partitions offrent une aide à la factorisation dans
les extensions.
Comme nous l'avons noté plus haut, le polynôme $f$ est une $H$-résolvante
avec $H=S_1\times S_{n-1}$. Supposons qu'on cherche à factoriser un polynôme de
$k[x]$ dans une
extension $k^\prime$ de $k$. Pour tout groupe de Galois possible $G$
de $f$ sur $k$, on détermine, en fonction de $G$, 
le groupe de Galois $\mathcal G$ qu'aurait $f$ dans 
l'extension $k^\prime$. Par exemple, pour $k^\prime=k(\alpha_1)$, 
${\mathcal G}=$Stab$(G,1)$. Alors $P({\mathcal G},H)$ est la liste des
degrés possibles des facteurs irréductibles de $f$ sur
$k^\prime$. Il est alors possible d'établir une table excluant des
types de factorisations de $f$ dans $k^\prime$.

Enfin, J.L. Lagrange a introduit les résolvantes pour généraliser la
résolution par radicaux. Il avait également introduit la fameuse
résolvante de Vandermond-Lagrange dont l'invariant est 
$$
x_1+\epsilon x_2 + \cdots + \epsilon^{n-1} x_n \quad .
$$
Ses résolvantes ont été utilisées plus de deux siècles plus tard dans
la résolution par radicaux des 
polynômes résolubles de degré 5 (voir \cite{MR1079014}) et de degré 6 
(voir \cite{MR1793923}). Dans ce dernier article, l'auteur appelle
``Galois resolvent'' ce qui en fait est la résolvante de
Lagrange. Il utilise la même méthode que dans \cite{LITP9361} (voir
aussi \cite{ArnaudiesValibouze:97}) pour déterminer une sous-matrice
de la matrice des partitions relative à $S_6$ (i.e. il calcule les cardinaux
des orbites).


\end{document}